# A New *Lecture-Tutorial* for Teaching Interferometry to Astro 101 Students

**Colin S. Wallace,** University of North Carolina at Chapel Hill, Chapel Hill, NC
**Chase Hatcher,** Cannon School, Concord, NC
**Timothy G. Chambers,** University of Michigan, Ann Arbor, MI
**Seth D. Hornstein,** University of Colorado Boulder, Boulder, CO
**Julia Kamenetzky,** Westminster College, Salt Lake City, UT
**Edward E. Prather,** University of Arizona, Tucson, AZ

The ground-breaking image of a black hole's event horizon, which captured the public's attention and imagination in April 2019, was captured using the power of interferometry: many separate telescopes working together to observe the cosmos in incredible detail. Many recent astrophysical discoveries that have revolutionized the scientific community's understanding of the cosmos were made by interferometers such as LIGO, ALMA, and the Event Horizon Telescope. Astro 101 instructors who want their students to learn the science behind these discoveries must teach about interferometry. Decades of research show that using active learning strategies can significantly increase students' learning and reduces achievement gaps between different demographic groups over what is achieved from traditional lecture-based instruction.[1] As part of an effort to create active learning materials on interferometry, we developed and tested a new *Lecture-Tutorial* to help Astro 101 students learn about key properties of astronomical interferometers. This paper describes this new *Lecture-Tutorial* and presents evidence for its effectiveness from a study conducted with 266 Astro 101 students at the University of North Carolina at Chapel Hill.

  A *Lecture-Tutorial* is a pencil and paper activity containing carefully sequenced Socratic-style questions that guide students toward more expert-like understandings of a specific topic. *Lecture-Tutorials* are designed to be used during class, but after students have received a short, introductory lecture on the associated topic. *Lecture-Tutorials* engage learners with one or more representations, including graphs, tables of data, and pictures and are designed to foster conversations between students in which they must discuss and defend their reasoning about a given topic. Consequently, *Lecture-Tutorials* are most effective when they are done by collaborative groups of 3-4 students.[2] Multiple studies show that *Lecture-Tutorials* significantly improve students' conceptual understandings and reasoning abilities.[3]

  The new *Lecture-Tutorial*, which is titled "Properties of Astronomical Interferometers," focuses on two properties: *resolution* and *light-collecting power* (LCP). We acknowledge that these are not the only interferometer properties that one could teach. There are many other physics principles and engineering details that are necessary for a complete and comprehensive understanding of how interferometers work. We restricted our attention to resolution and LCP for two reasons. First, Astro 101 instructors often introduce resolution and LCP when discussing the important properties of single-aperture telescopes. Students that have this background can appreciate and understand the advantages of an array of telescopes. Second, resolution and LCP are conceptually rich topics, but they are also tractable enough that they can be addressed within the page limitations of a *Lecture-Tutorial*. The combination of variables presented to address resolution and LCP allows students to engage in the type of quantitative reasoning that is often an explicit learning objective of introductory-level, general education science courses. The "Properties of Astronomical Interferometers" *Lecture-Tutorial* helps students understand how resolution depends on the observed wavelength and the interferometer's baseline and how LCP depends on the number of telescopes and the average surface area of each telescope.

**Overview of the "Properties of Astronomical Interferometers" *Lecture-Tutorial***

At the beginning of the *Lecture-Tutorial*, students are shown a graph on which the vertical axis corresponds to LCP and the horizontal axis corresponds to the number of telescopes in an interferometer. Three points are shown on the graph. Each point represents a different interferometer. This graph is shown in Figure 1. In this figure, two interferometers are intentionally chosen to have the same LCP while two are chosen to have the same number of telescopes. This explicitly makes students consider the effect that size and number of telescopes each have on total LCP, both independently and together. Students must use this graph to rank the interferometers based on their LCPs and to determine which of the interferometers has the smallest telescopes.

[Figure 1 about here.]

Figure 2 shows the next graph that students must interpret. Once again, each dot represents a different interferometer. The vertical axis now corresponds to the observed wavelength and the horizontal axis corresponds to the baseline. Notice that two interferometers operate at the same wavelength while two have the same baseline. This gives students the opportunity to reason about how the observed wavelength and baseline affect an interferometer's resolution. Students must rank the interferometers based on the value for the resolution and then determine which interferometer could produce images with the smallest details.

[Figure 2 about here.]

We found that we had to be especially careful with the language we used to describe resolution. When calculating the resolution, one must divide the observed wavelength by the interferometer's baseline. The smaller the resulting number, the more fine details the interferometer can detect. This is opposite to how resolution is used in a non-scientific usage (e.g. HDTV). Students think that a large (or high) calculated value for resolution is required to see fine details. In order to be precise with our language, the *Lecture-Tutorial* always refers to the resolution as a number or value and in our lecture preceding the *Lecture-Tutorial* we emphasize that this number must be small if you want to detect fine details.

The *Lecture-Tutorial* then presents students with four images of the same galaxy (Figure 3). Students are told that each image was made by a different interferometer. Students then have to determine which images were made by interferometers with small resolution values and which were made by interferometers with large resolution values. Students also have to determine which images were made by interferometers with a large LCP and which were made by interferometers with a small LCP.

[Figure 3 about here.]

Two images clearly show smaller details while two images show dimmer features. Note that the brightest part of the galaxy (represented in white) has the same brightness in all four images. This is not a mistake. In radio astronomy, increasing the LCP does not make the observed object appear brighter. This is because the astronomical signal is often extremely weak and almost everything that is detected is noise. Increasing the LCP helps decrease the noise and therefore increase the signal-to-noise ratio, allowing us to detect dimmer objects and features, such as the dimmer, grey arms that appear in two of the images of the galaxy. This is one way in which radio astronomy is different from optical astronomy, where increasing the LCP will make the observed object appear brighter.

[Figure 4 about here.]

Next, students are shown four different interferometers (Figure 4). The interferometers differ in the number of telescopes, the size of each telescope, the baseline, and the wavelength at which they are observing. Students must evaluate the range of variables presented by these interferometers in order to rank the interferometers based on their LCP. Next, students rank the interferometers based on their resolution. Students are then asked to match each interferometer with the image of the galaxy (from Figure 3) it could have produced. This set of tasks requires students to coordinate all of the disciplinary affordances of Figures 3 and 4 and serves as a critical step in their development of a coherent explanatory model for understanding interferometers.

[Figure 5 about here.]

Figure 5 shows the final task students must complete. They are given representations of four different interferometers, all of which are observing the same wavelength of light, and must assign each interferometer to one of four teams of astronomers. Each team wants to observe a different object and the resolution and LCP requirements for that object are given. This task requires students to use everything they have learned about resolution and LCP in order to evaluate which interferometer should be assigned to which team.

**Assessing the *Lecture-Tutorial*'s effectiveness**
We developed a ten-item multiple-choice test to assess students' understandings of the concepts addressed by the new *Lecture-Tutorial*. We administered the test to students taking Astro 101 during fall 2016 and spring 2017 at the University of North Carolina at Chapel Hill. The test was administered twice each semester: before students received any instruction on interferometry and after they received a lecture, with Think-Pair-Share voting questions, and completed the *Lecture-Tutorial*. Students recorded their answers on multiple-choice bubble forms. We removed from our analysis all forms in which the responses indicated that the assessment was not taken seriously (e.g. entirely blank or repeated answers and other obviously recognizable patterns). We then removed all responses from students who took only the pre-test or post-test. This left us with a data set of 121 matched pre- and post-test responses for fall 2016 and 145 for the spring 2017.

[Table 1 about here.]

Table 1 shows the item difficulty (percentage correct) for each item, pre- and post-instruction, for both semesters. It also shows the discrimination of each item, which is a measure of the correlation between students' scores on that item and their total score on the test. We are only showing the discriminations for the post-instruction scores; the discrimination values for the pre-test were all very low, which is to be expected since many students may be guessing the answers to each item. There are several things to notice about the data in this table. First, students' pre-test performance varied widely from item to item. On some items, only 8% of students selected the correct answer, while on others 60% were correct. There were also three items (4, 8, and 10) on which students from the fall 2016 class outperformed students from the spring 2017 on the pre-test. We do not completely understand why the two semesters performed differently on these three items. Since the post-test results were similar on these items, more students in Fall 2016 may have been better prepared for these items by instruction or experience prior to the course.

Any differences between the two classes disappeared by the time they took the post-test. Overall, students did quite well. The percentage correct on each item ranged from 63% to 89%. This suggests that many students had increased their understandings of the content addressed by the *Lecture-Tutorial* by the time they took the post-test. We will address this point in more detail below.

The post-test item difficulties and discriminations shown in Table 1 all fall within conventionally accepted limits, with the sole exception of the discrimination for item 10 for the fall 2016.[4] We are not sure why this discrimination value was so low. However, every other calculated difficulty and discrimination value suggests that the items were functioning properly.

We also use each semester's post-test data to calculate Cronbach's $\alpha$.[5] Cronbach's $\alpha$, a measure of reliability or internal consistency of a set of data, compares the variance of test scores as a whole to the variance of individual items on the test so that it is high (approaching 1) when covariances among items are high. When applied to test scores, $\alpha$ essentially tells us how likely we would be to get similar results if we gave the same test to different, but representative, populations. Our average $\alpha$ between the two semesters is 0.675, which is just below the usual mark for acceptability (0.7).[6] We argue that our $\alpha$ is still acceptable because the value of $\alpha$ is sensitive to test length and could be lowered by this shorter test. Furthermore, it is shortsighted to treat an arbitrary cutoff value in a continuous variable as an absolute requirement since other factors like meaningful content coverage and validity are arguably as, or more, important for assessing learning.[7]

Having established that the test appears to be reliable and that the items appear to be functioning properly, we now present additional evidence that students' understandings significantly increased pre- to post-test. Figures 6 and 7 show the distribution of students' scores pre- and post-instruction, respectively. The mean, median, and mode of the post-instruction distribution are all substantially larger than they are for the pre-instruction distribution, regardless of semester.

[Figure 6 about here.]
[Figure 7 about here.]

We also calculated Hake's normalized learning gain $\langle g \rangle$ as well as Marx and Cumming's normalized change $c$.[8] For the fall 2016, $\langle g \rangle = 0.71$ and $c = 0.64$. For the spring 2017, $\langle g \rangle = 0.71$ and $c = 0.69$. These values are all quite high – in fact, both values of $\langle g \rangle$ fall within the "high-$\langle g \rangle$" region defined by Hake. The values of the normalized learning gain and normalized change, along with the distribution of scores in Figures 6 and 7 and the item difficulties in Table 1, all strongly imply that the new *Lecture-Tutorial*, plus the associated interactive lecture on interferometry that we developed, can be used as part of an Astro 101 class to significantly improve students' understandings of resolution and LCP in the context of astronomical interferometers.

**Conclusions**
We have developed a new *Lecture-Tutorial*, called "Properties of Astronomical Interferometers." This *Lecture-Tutorial* helps students understand how resolution depends on the observed wavelength and the interferometer's baseline and how LCP depends the number of telescopes and the average surface area of each telescope's mirror/dish. To test the efficacy of this *Lecture-Tutorial*, we developed a ten-item multiple choice test, which we administered pre- and post-instruction to Astro 101 students in two different semesters. An analysis of multiple classical test theory statistics (item difficulty, item discrimination, and Cronbach's $\alpha$) revealed that the test is reliable and the items are functioning properly. Students show dramatic improvements on the test, pre- to post-instruction. We conclude that the new *Lecture-Tutorial* can be an effective addition to an Astro 101 class to improve students' understanding of how astronomers use interferometry in their study of the universe.

**Availability of the *Lecture-Tutorial***
A copy of the *Lecture-Tutorial*, as well as the accompanying lecture slides and Think-Pair-Share questions, can be found at the website for the Center for Astronomy Education: LINK.

**Acknowledgements**
This work was made possible by the generous support of Associated Universities, Inc. (AUI) and NSF AAPF award AST-1402193. Additionally, we would like to thank Alice Churukian for her help with the data collection, as well as the students who participated in this study.

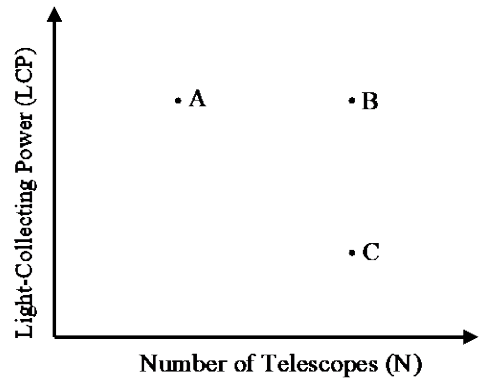

The Light-Collecting Power (LCP) and Number of Telescopes (N) of three interferometers (A, B and C) are plotted in the graph at right.

1) Rank the light-collecting power (LCP) of the interferometers (A, B and C) from greatest to least.

2) Which of the interferometers has the smallest - sized telescopes? Explain your reasoning.

Figure 1. A graph of light-collecting power (LCP) vs. number of telescopes. The three points (A-C) represent three different interferometers. Students must use this graph to answer the questions shown to the left of the graph.

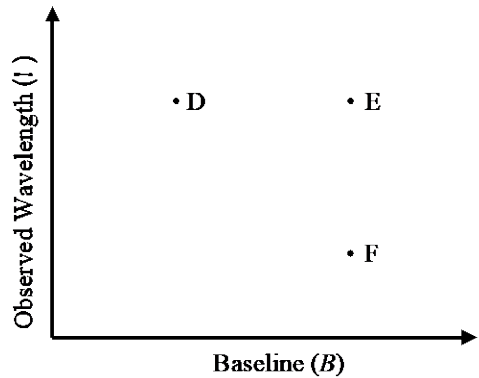

The Observed Wavelength ($\lambda$) and Baseline ($B$) of three interferometers (D, E and F) are plotted in the graph at right.

3) Rank the resolution (R) of the three interferometers (D, E and F) from smallest to largest number.

4) If all three interferometers (D, E, and F) observe the same object, which interferometer could produce images with the smallest details? Explain your reasoning.

Figure 2. A graph of observed wavelength vs. baseline. The three points (D-F) represent three different interferometers. Students must use this graph to answer the questions shown to the left of the graph.

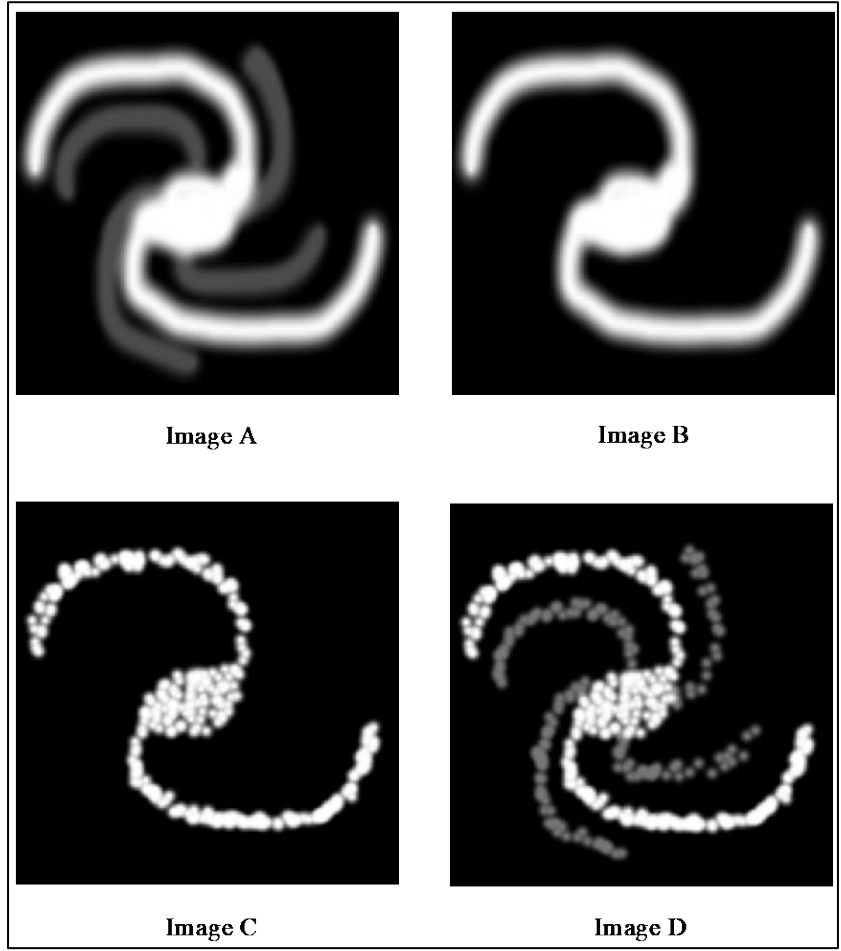

**Figure 3. Four different simulated images (A-D) of the same galaxy. Students must determine which images were made by interferometers with large vs. small LCP, as well as which were made by interferometers with large vs. small resolution values.**

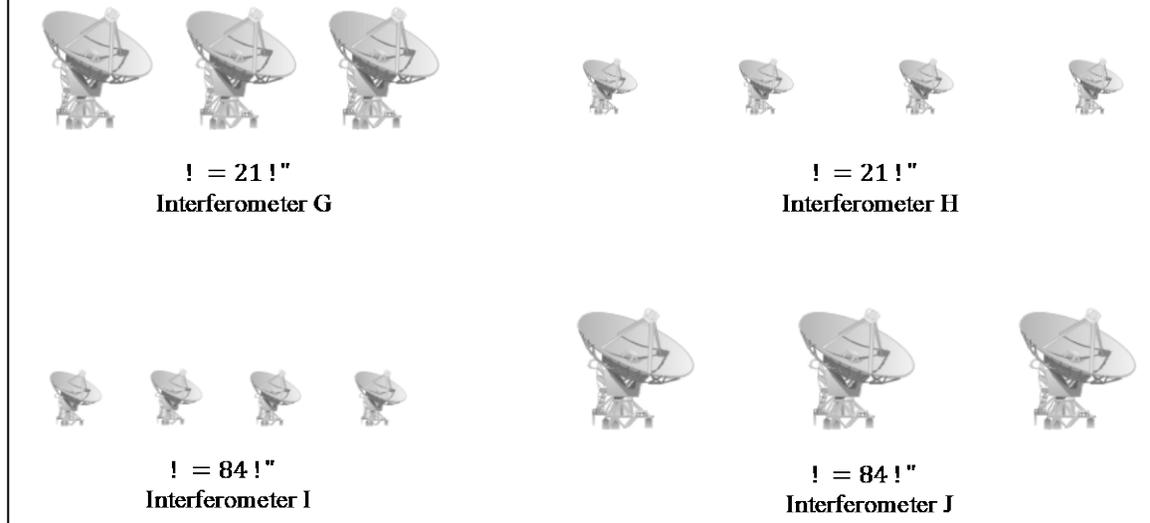

Figure 2, below, depicts four different interferometers. The label for each interferometer includes the wavelength at which it observes the sky. The area of each large telescope is 4 times the area of each small telescope and the baseline of Interferometers H and J is twice the baseline of Interferometers G and I.

$\lambda = 21\lambda"$
Interferometer G

$\lambda = 21\lambda"$
Interferometer H

$\lambda = 84\lambda"$
Interferometer I

$\lambda = 84\lambda"$
Interferometer J

**Figure 4. Four different interferometers. Students must match each interferometer with the image it produced in Figure 3.**

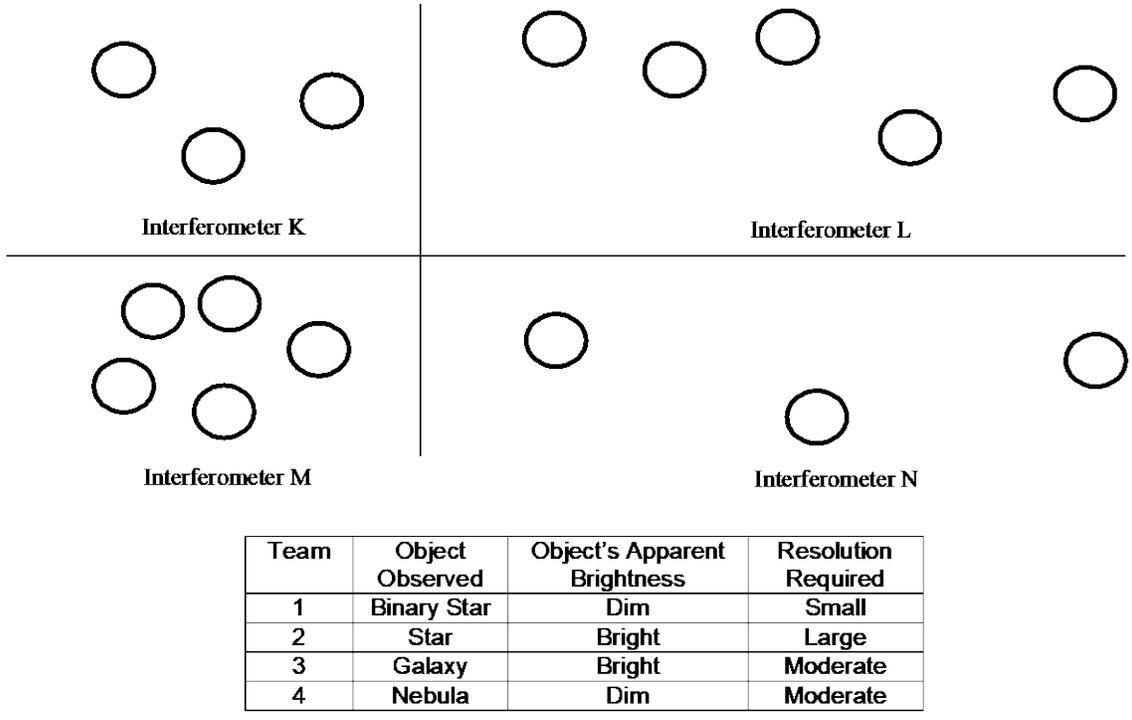

Figure 5. For the final task in the *Lecture-Tutorial*, students are presented with these four different interferometers (K-N) and the observing needs of four teams (1-4) of astronomers. Students must assign each interferometer to the appropriate team.

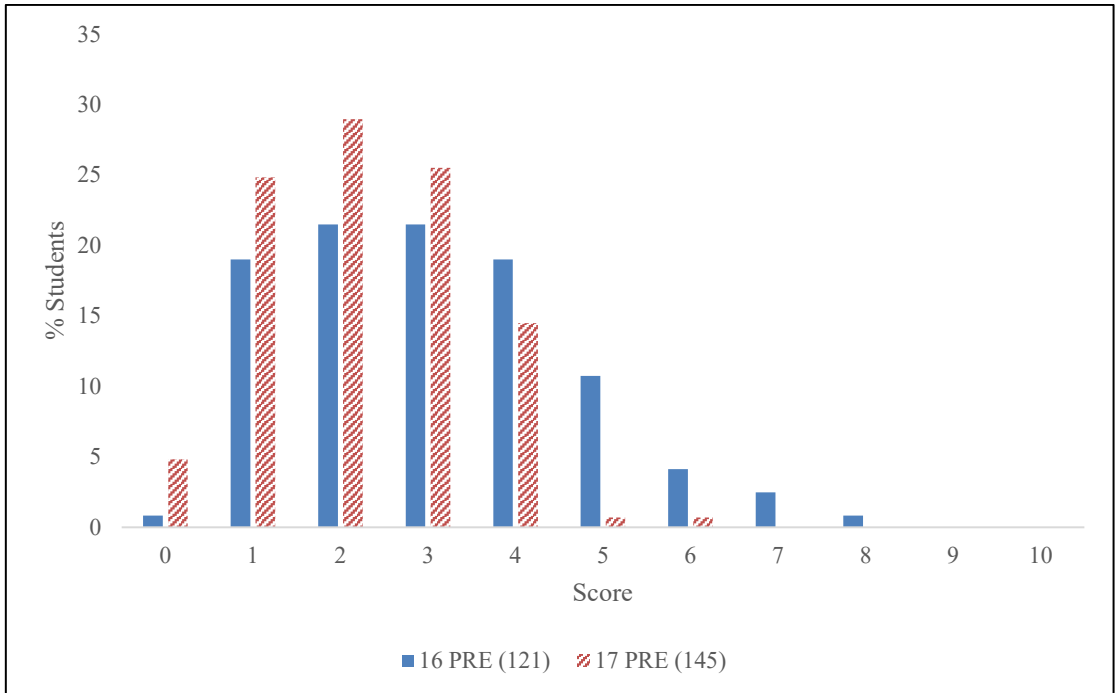

Figure 6. The distribution of total pre-test scores for both the fall 2016 and spring 2017 semesters. For the fall 2016, the distribution's mean is 3.1, the median is 3.0, and the mode is 3.0. For the spring 2017, the distribution's mean is 2.3, the median is 2.0, and the mode is 2.0.

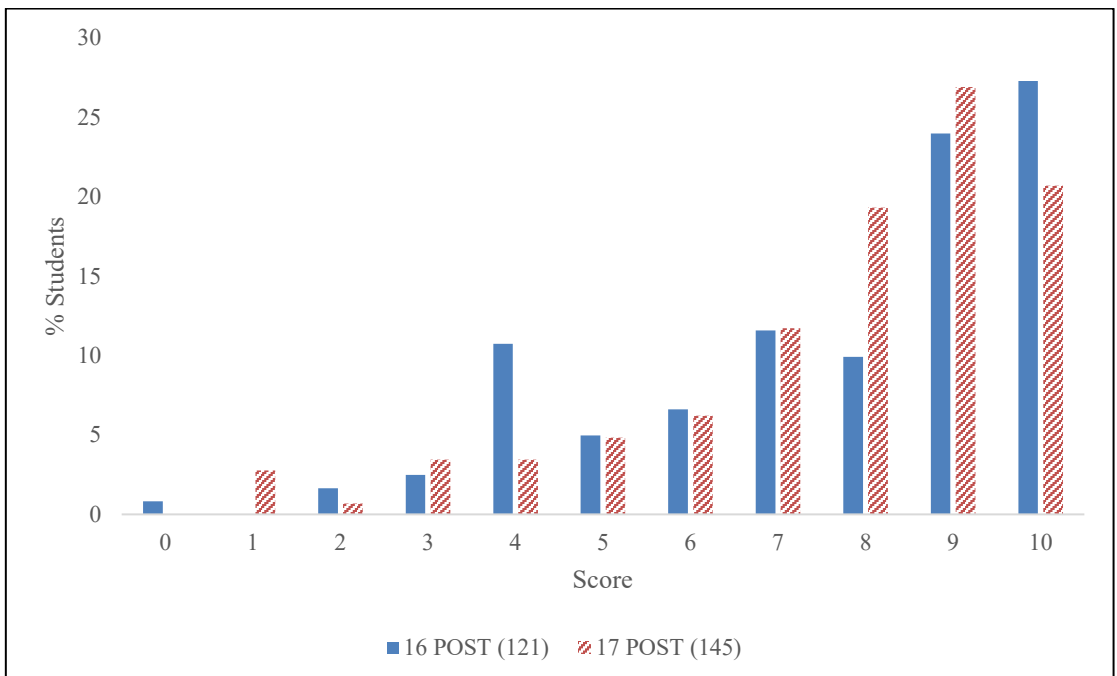

Figure 7. The distribution of total post-test scores for both the fall 2016 and spring 2017 semesters. For the fall 2016, the distribution's mean is 7.7, the median is 9.0, and the mode is 10.0. For the spring 2017, the distribution's mean is 7.8, the median is 8.0, and the mode is 9.0.

|      | Item Difficulty |          |          |          | Item Discrimination |          |
| ---- | --------------- | -------- | -------- | -------- | ------------------- | -------- |
| Item | '16 pre         | '17 pre  | '16 post | '17 post | '16 post            | '17 post |
| 1    | 0.08            | 0.08     | 0.67     | 0.71     | 0.47                | 0.40     |
| 2    | 0.31            | 0.25     | 0.77     | 0.83     | 0.39                | 0.34     |
| 3    | 0.40            | 0.47     | 0.68     | 0.63     | 0.33                | 0.40     |
| 4    | 0.40            | 0.14     | 0.75     | 0.68     | 0.45                | 0.30     |
| 5    | 0.26            | 0.18     | 0.79     | 0.77     | 0.44                | 0.46     |
| 6    | 0.12            | 0.12     | 0.79     | 0.83     | 0.53                | 0.40     |
| 7    | 0.43            | 0.39     | 0.79     | 0.82     | 0.60                | 0.48     |
| 8    | 0.60            | 0.46     | 0.89     | 0.89     | 0.36                | 0.46     |
| 9    | 0.17            | 0.10     | 0.80     | 0.87     | 0.62                | 0.37     |
| 10   | 0.28            | 0.08     | 0.74     | 0.72     | 0.10                | 0.33     |

**Table 1. The pre- and post-test item difficulties and the post-test item discriminations for the fall 2016 and spring 2017 data sets.**